\documentclass[aps,prb,twocolumn,showpacs,amsmath,amssymb]{revtex4-1}
\usepackage{txfonts}
\usepackage{graphicx}
\usepackage{cases}
\usepackage{breqn}
\usepackage{amsmath}
\usepackage{blkarray}
\usepackage{dcolumn}
\usepackage{bm}
\usepackage{multirow}
\usepackage{float}
\usepackage{tabularx}
\usepackage{color}
\usepackage[dvipsnames]{xcolor}
\usepackage{hyperref}
\usepackage{graphicx}
\usepackage{caption}
\usepackage{subcaption}
\usepackage[export]{adjustbox}
\def\eV{\,\textrm{eV}}

\newcommand{\be}{\begin{equation}}
\newcommand{\ee}{\end{equation}}
\newcommand{\bea}{\begin{eqnarray}}
\newcommand{\eea}{\end{eqnarray}}
\newcommand{\balg}{\begin{align}}
\newcommand{\ealg}{\end{align}}

\begin{document}

%%%%%%%%%%%%%%%%%%%%%%%%%%%%%%%%%%%%%%%%%%%%%%%%%%%%%%%%%%%%%%%%%%%%%%%%%%%%%%%%%%%%%%

\title{ cRPA calculation of on-site and nearest neighbor Coulomb interaction of $LaNiO_2$ } 

\author{Long Zhang$^{1}$ } 
%\author{Urs R. H{\"a}hner$^{2}$ } 
%\author{Thomas Schulthess$^{2}$ }
\author{Hai-Ping Cheng$^{1}$ }
%  \email{Contact Email: cheng@qtp.ufl.edu; Tel: (352) 392 6256 \\ }

\affiliation{$^{1}$ Department of Physics and The Quantum Theory Project$,$  University of Florida$,$  Gainesville FL 32611$,$ USA}
             
%%%%%%%%%%%%%%%%%%%%%%%%%%%%%%%%%%%%%%%%%%%%%%%%%%%%%%%%%%%%%%%%%%%%%%%%%%%%%%%%%%%%%%

\begin{abstract}
We present first-principle calculation of the on-site and nearest neighbor Coulomb interaction strength of the Ni $d$ orbitals in bulk $LaNiO_{2}$, using the constrained Random Phase Approximation method. The nearest neighbor correlation within Ni-O plane turns out to be more significant when considering the frequency dependent $U(\omega)$, which can be as strong as about 25\% of the on-site value at medium and high frequencies. The inter Ni-O plane nearest neighbor correlation is found to be the same strength as that within the Ni-O plane, indicating the material is non-locally correlated also between the Ni-O planes. 
\end{abstract}

\maketitle

\section{Introduction} 

Since the discovery of high-temperature superconductivity in the cuprates \cite{BaLaCuO_1986}, tremendous theoretical and experimental efforts have been devoted to the physics of this family of materials \cite{nature14165,RevModPhys.78.17,RevModPhys.84.1383,RevModPhys.72.969,RevModPhys.75.473}. 
Searching for the cuprate-like superconductivity candidates in the family of transition metal oxides is one of the research directions in this area. By studying the similarities and differences, theorists hope to understand the mechanism and expand the utility of superconductivity \cite{Norman_2016} in functional materials. 
Although a complete and unambiguous understanding of its nature has not been reached, some essential features for superconductivity have been highlighted for searching for cuprate-like materials. 
The main common features include the two-dimensional electronic structure and magnetism \cite{RevModPhys.84.1383}, antiferromagnetically interacting S=1/2 moments, substantial d–p hybridization, and the large orbital polarization (one-band physics). 
Nickelates have been studied along this line as a candidate of non-Cu-based but cuprate-like superconductor \cite{Zhang2017,science.1202647,PhysRevLett.107.206804,PhysRevLett.103.016401,PhysRevLett.106.027001,PhysRevLett.100.016404}. 
One important milestone in this direction is the discovery of superconductivity at 9-15K in the infinite-layer nickelate $Nd_{1-x}Sr_{x}NiO_{2}$ grown on $SrTiO_{3}$ substrate \cite{Li2019,PhysRevLett.125.027001,PhysRevLett.125.147003}. 
The mother compound is $NdNiO_{2}$ which was synthesized about two decades ago \cite{HAYWARD2003839} and the closely related compound $LaNiO_{2}$ was synthesized much eariler \cite{F29837901181_LaNiO2_1983}. 
Early theoretical studies \cite{PhysRevB.70.165109} had excluded $LaNiO_{2}$ from the cuprate analogs because the d-p hybridization is weak and the Fermi surface is non-cuprate like, while recent discover of $Nd_{1-x}Sr_{x}NiO_{2}$ brought the properties of both $NdNiO_{2}$ and $LaNiO_{2}$ back to attention \cite{Zhang2021_LaNiO2_NdNiO2}. 

$NdNiO_{2}$ and $LaNiO_{2}$ are isostructural to the infinite-layer cuprates with a flat $NiO_{2}$ plane of the square lattice of monovalent $Ni^{1+}$ cations. 
The $Ni^{1+}$ cation has one hole in the $d_{x^{2}-y^{2}}$ orbital and possesses the same $3d^{9}$ electron configuration counting as $Cu^{2+}$ cations in the undoped cuprates. 
Superconductivity in bulk $NdNiO_{2}$ has not been observed \cite{PhysRevMaterials.4.084409}. 
Pristine $NdNiO_{2}$ and $LaNiO_{2}$ have metallic behavior with no sign of long-range magnetic order down to low temperatures \cite{PhysRevMaterials.4.084409,HAYWARD2003839,doi:10.1021/ja991573i}, suggesting a weak or mediate correlation effect. 
It has been pointed out that the effect of $Sr$ doping gives a more pure single $d_{x^{2}-y^{2}}$ band cuprate-like picture \cite{PhysRevX.10.011024}, which makes the value of the Coulomb interaction strength of the $d_{x^{2}-y^{2}}$ orbital an important quantity to characterize the system. 
Indeed the on-site Coulomb interaction $U$ was included in all theoretical and first-principle studies, and the value of on-site $U$ are mostly empirical in the range of 3-8 eV. 
A dedicated calculation of $U$ from first principle is desirable for not only the on-site correlation but also the possible non-local correlation. 
It was pointed out that \cite{PhysRevResearch.3.033157} the nearest neighbor Coulomb interaction in cuprates affects the stability of superconductivity and the phase competition among various phases. 
While dedicated studies on both on-site and nearest neighbor Coulomb interactions are still missing for the newly discovered Nickelate materials. We are thus motivated to calculate the on-site and the various nearest neighbor Coulomb interactions from first principle for $LaNiO_{2}$. 

Previous study \cite{PhysRevLett.125.077003} had shown that $LaNiO_{2}$ and $NdNiO_{2}$ give essentially the same band structure except for the Nd-4$f$ bands. And, studying $LaNiO_{2}$ instead of $NdNiO_{2}$ helps to avoid the issue of Nd-4$f$ \cite{PhysRevB.102.161118}. 
Therefore we study $LaNiO_{2}$ and suggest the calculated Coulomb interaction of Ni sites would be close to that of Ni in $NdNiO_{2}$, because neither Nd $f$-like bands nor La $f$-like bands are close to the Fermi level in DFT studies of these two materials and the method we used only considers the $d$-like bands of Ni. 

\begin{figure}[H]
  \includegraphics[width=0.9\columnwidth]{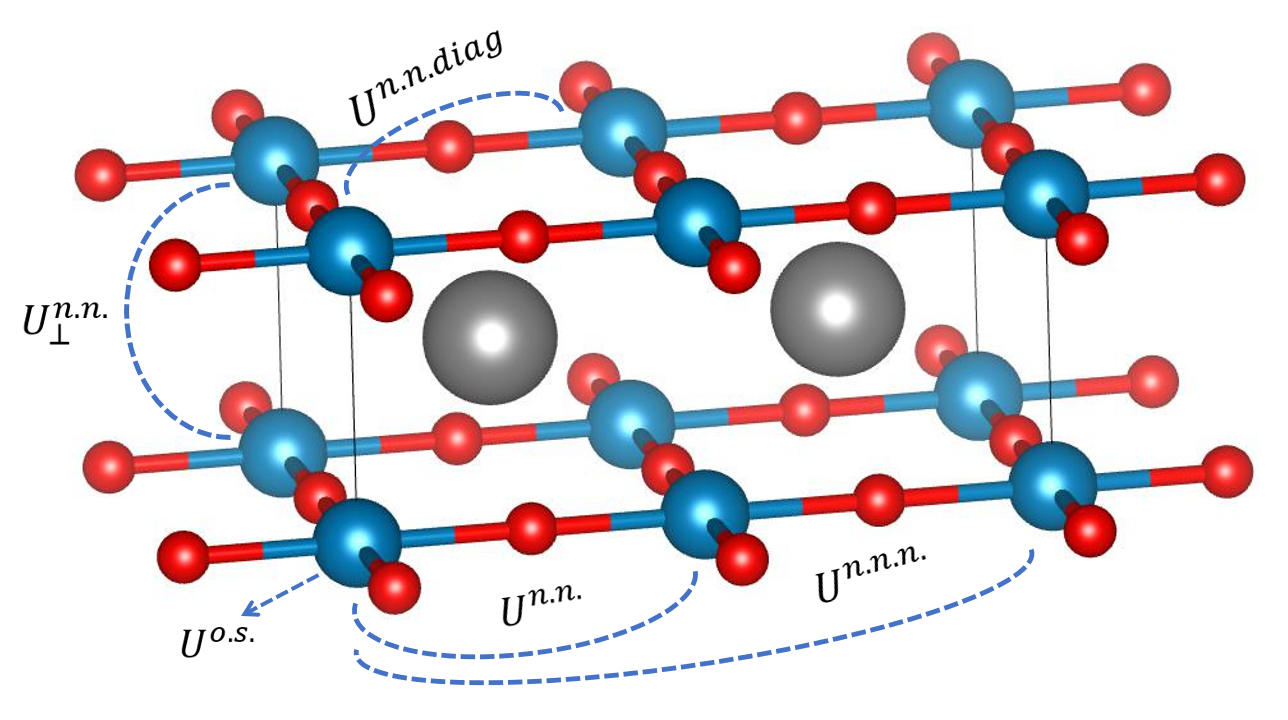}
  \caption
  {Crystal $LaNiO_{2}$ is displayed in a 2x1x1 supercell setup, in order to identify the various in-plane Coulomb interaction between the Ni atoms (blue balls): on-site (o.s.), nearest neighbor (n.n.), nearest neighbor in diagonal (n.n.diag) and next nearest neighbor (n.n.n.). The inter-plane nearest neighbor interaction is tagged with subscript "$\perp$". Red balls are Oxygen atoms. Gray balls are Lanthanum atoms.}
  \label{crystal_LaNiO2}
\end{figure}

In the current work, we consider the following Coulomb interactions in $LaNiO_{2}$, within Ni-O plane: the on-site $U^{o.s.}$, the nearest neighbor $U^{n.n.}$, the nearest neighbor in diagonal $U^{n.n.diag}$, the next nearest neighbor $U^{n.n.n.}$, and between two Ni in adjacent Ni-O planes: $U_{\perp}^{n.n.}$. These have been shown in Fig.\ref{crystal_LaNiO2}.

% --- figure 

$\mathit{Outline:}$ The remainder of the paper is organized as follows. Section \textcolor{blue}{II} introduces the calculation methods, including the DFT and DFT results which is the base for the following constrained RPA (cRPA) calculation of the Coulomb interaction $U$. Section \textcolor{blue}{III} has the resulting $U$ matrices and the frequency dependence of several key matrix elements. Section \textcolor{blue}{IV} provides summary and conclusion.

\section{Calculation Methods}

In this section we first present the DFT band structure of $LaNiO_{2}$ and the orbital characters of the bands close to Fermi level that motivated the selection of the correlation window for subsequent model construction and cRPA calculation. 
Then we briefly go over the theory of cRPA. The resulting $U(\omega)$ is presented in the next section. 

\subsection{ DFT calculation }

The DFT calculation is done using the FP-LAPW method, as implemented in a modified version of the ELK code \cite{Anton_IEEE_paper_2010}. The ground state is calculated within the local density approximation (LDA). The muffin tin sphere radii are 2.2 $a_{0}$, 2.0 $a_{0}$ and 1.6 $a_{0}$ for La, Ni and O, respectively. A dense k-point grid of $16$x$16$x$16$ was used to perform Brillouin zone integration. The used lattice parameters for $LaNiO_{2}$ are $a=b=$ 3.96 {\AA} and $c=$ 3.37 {\AA}. $NdNiO_{2}$ has slightly different lattice parameters, $a=b=$ 3.92 {\AA} and $c=$ 3.28 {\AA}. As pointed out in \cite{PhysRevX.10.011024}, we also found that a non-magnetic calculation of $LaNiO_{2}$ at the lattice parameters of $NdNiO_{2}$ does not give rise to any important changes in the band structure. Thus we stick to the original $LaNiO_{2}$ lattice parameter in all calculations. 

% --- figure 
\begin{figure}[H]
  \includegraphics[width=1.0\columnwidth]{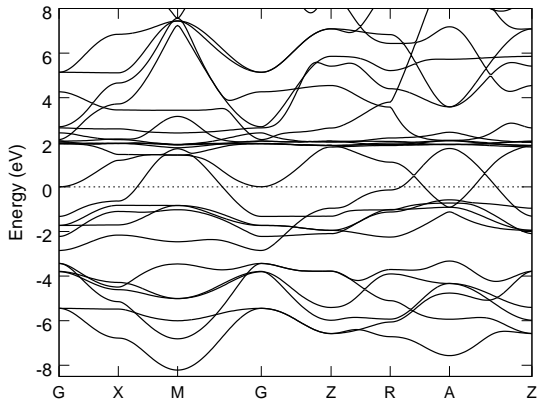}
  \caption
  {Non-magnetic ground state band structure of $LaNiO_{2}$. Fermi level is at zero.}
  %(solid lines), and band characters (open circles) calculated by projecting Bloch states onto atomic orbital states. The radius of the open circles are proportional to the weight of the atomic states. Fermi level is at zero.} 
  \label{scf_bands}
\end{figure}

\noindent The non-magnetic ground state band structure is shown in Fig.\ref{scf_bands}. There are five $d$-like bands around the Fermi level in $[-3.0,+2.0] \eV$, representing the partially filled $d$ states of Ni, giving the material a metallic ground state. Below them in the  $ [-8.0,-3.0] \eV $ range are six bands showing Oxygen $p$ orbital character. The Ni $d$-like bands and $p$-like bands are separated by a small gap. Above the Ni $d$-like bands, crowded at about $+2.0 \eV$, there are sever flat band of the La $f$ character. The La $d$-like bands are entangled with other bands above the Fermi level with most weights about $+2.0 \eV$. 

% --- figure 
\begin{figure}[H]
  \centering
  \begin{subfigure}[b]{0.49\columnwidth}
    \includegraphics[width=\linewidth]{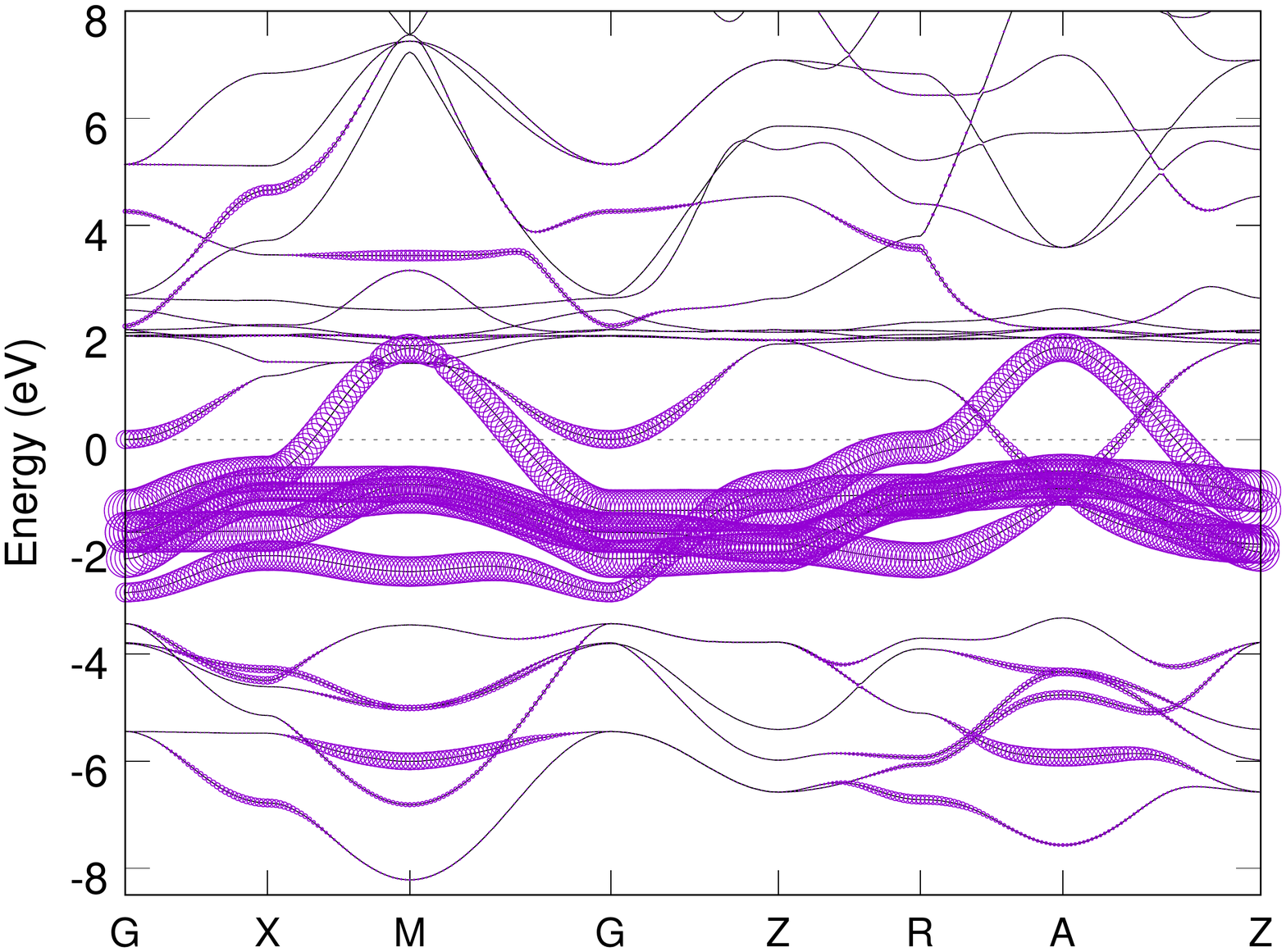}
     \caption{total Ni $d$}
  \end{subfigure}
  \begin{subfigure}[b]{0.49\columnwidth}
    \includegraphics[width=\linewidth]{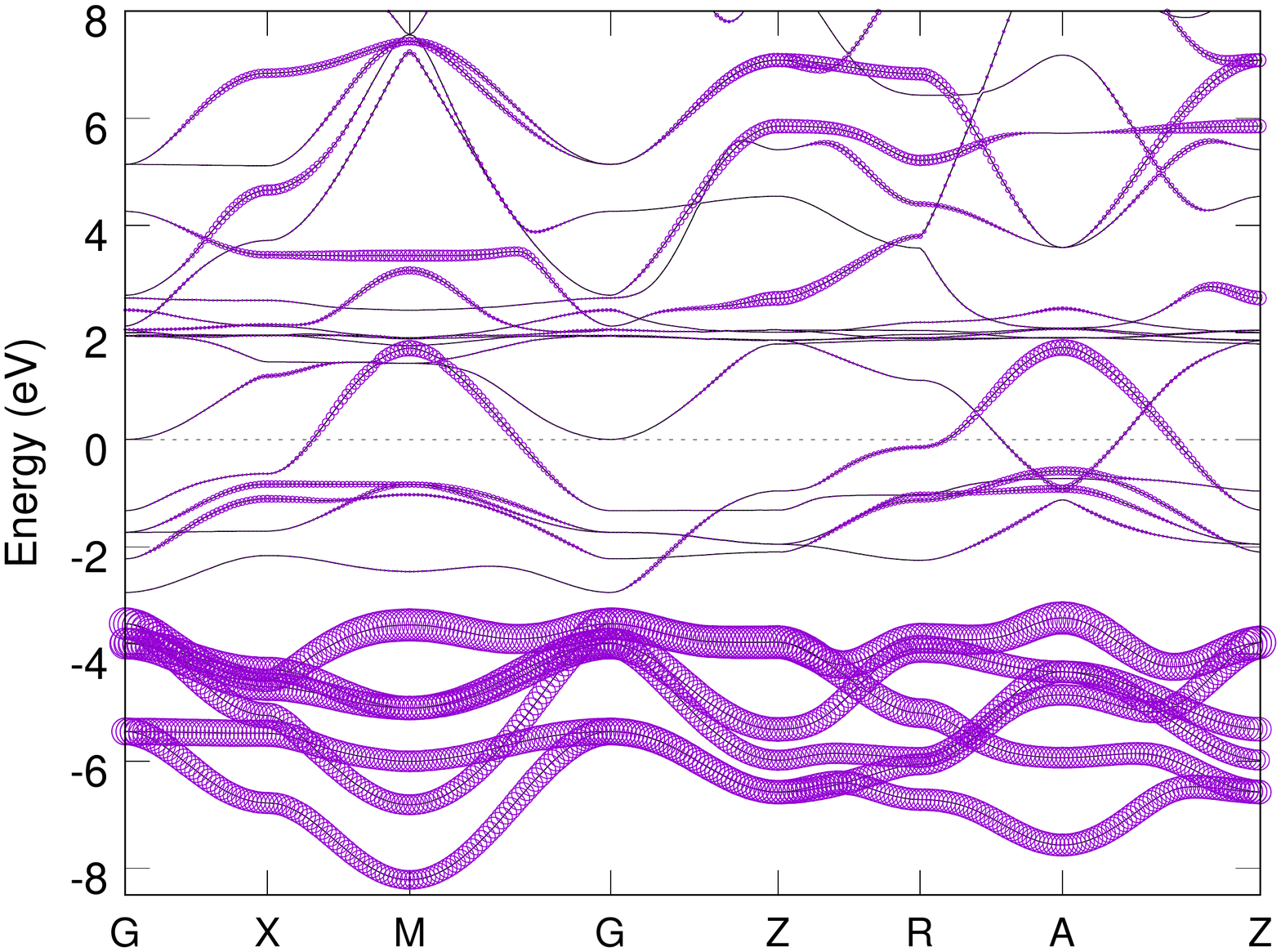}
    \caption{total O $p$}
  \end{subfigure}
  \begin{subfigure}[b]{0.49\columnwidth}
    \includegraphics[width=\linewidth]{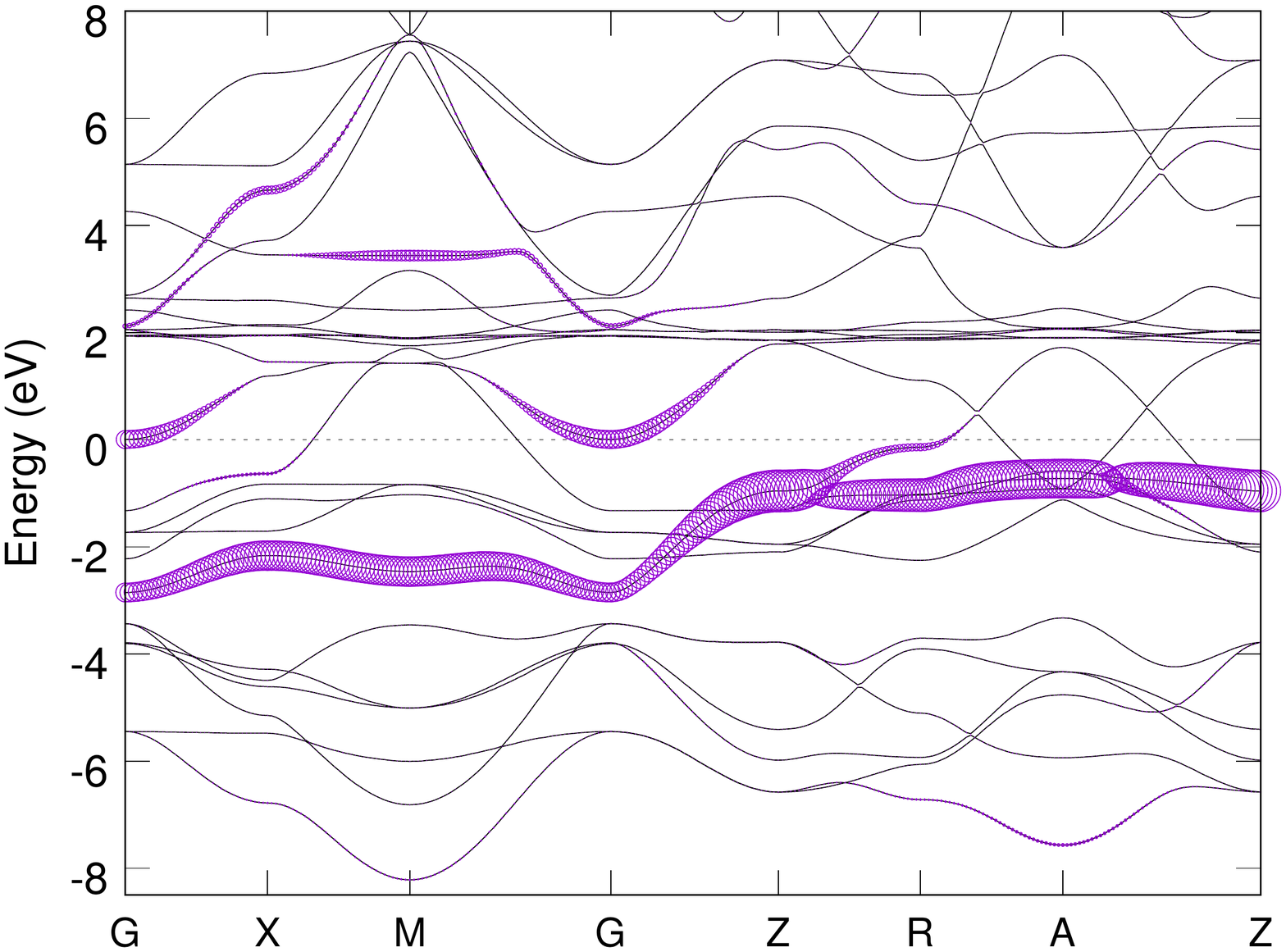}
    \caption{Ni $d_{3z^{2}-r^{2}}$}
  \end{subfigure}
  \begin{subfigure}[b]{0.49\columnwidth}
    \includegraphics[width=\linewidth]{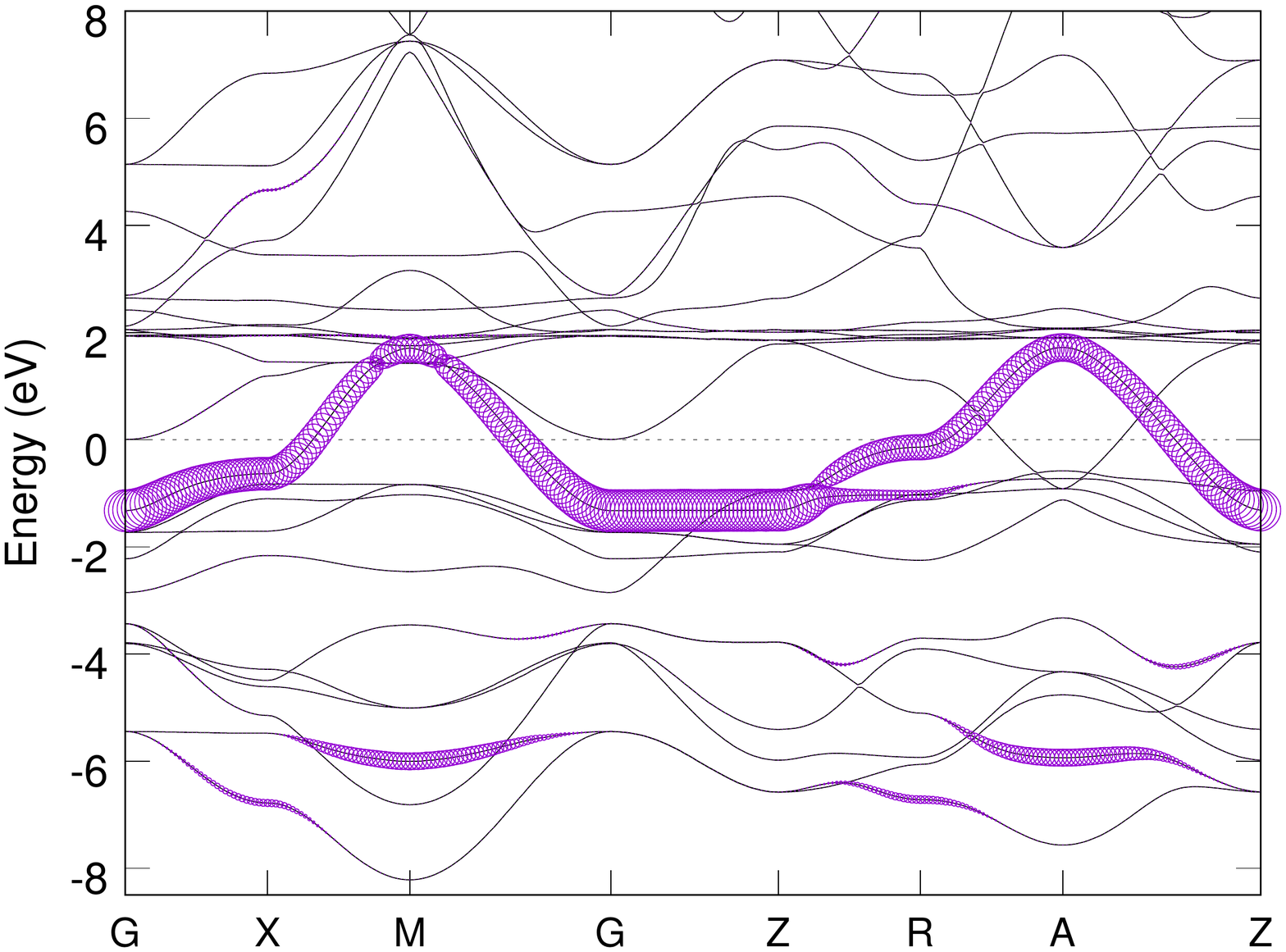}
    \caption{Ni $d_{x^{2}-y^{2}}$}
  \end{subfigure}
  \caption{Fat bands showing the amount of overlapping between Bloch states and Ni $d$ states and O $p$ states. (a) and (b): total $d$ and total $p$ character. (c) and (d): individual $d_{3z^{2}-r^{2}}$ and $d_{x^{2}-y^{2}}$ character.}
  \label{fat_bands_Ni}
\end{figure}
% --- figure 

The orbital character of the Bloch bands are often indicated by its overlapping with crystal field split states like the $t_{2g}$ and $e_{g}$ of the $d$ orbital for example. That helps with identifying proper energy windows for Wannier downfolding in the next step. The Ni $d$-like bands and O $p$-like bands are identified in the fat band plots in Fig.\ref{fat_bands_Ni}. It's clearly seen the Ni $d$ weights are in $[-3.0,+2.0] \eV$ and the $d$-$p$ mixing is not significant though the $p$ weights spread up to above the $f$-like bands of La. The individual Ni $d_{x^{2}-y^{2}}$ character dominants around the Fermi level and is mainly responsible for the physics properties. The Ni $d_{3z^{2}-r^{2}}$-like band is entangled with the Ni $d_{x^{2}-y^{2}}$ band in the $k$-path Z-R, which motivates a two-band model construction later. 

\begin{figure}[H]
  \centering
  \begin{subfigure}[b]{0.49\columnwidth}
    \includegraphics[width=\linewidth]{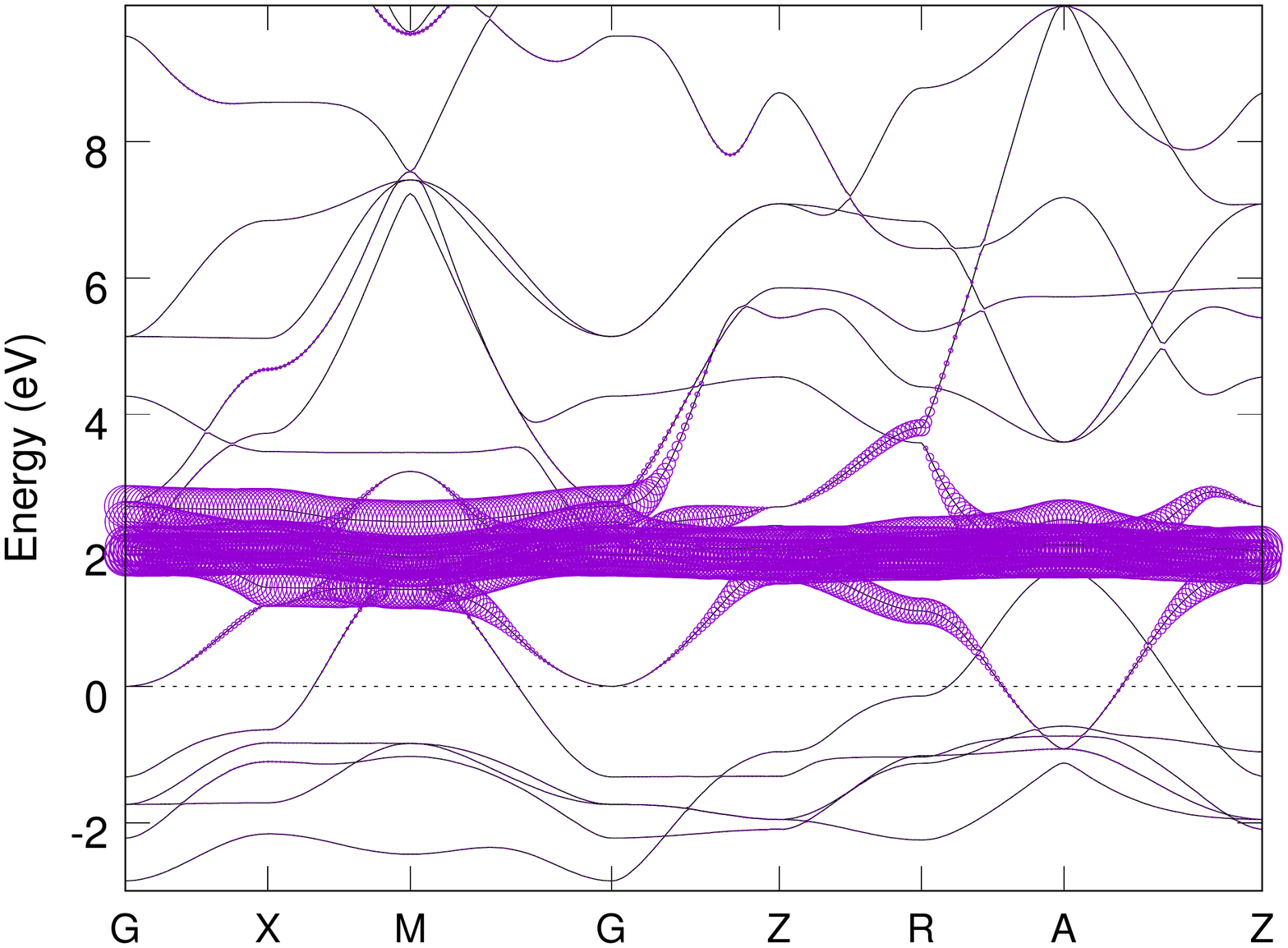}
     \caption{total La $f$}
  \end{subfigure}
  \begin{subfigure}[b]{0.49\columnwidth}
    \includegraphics[width=\linewidth]{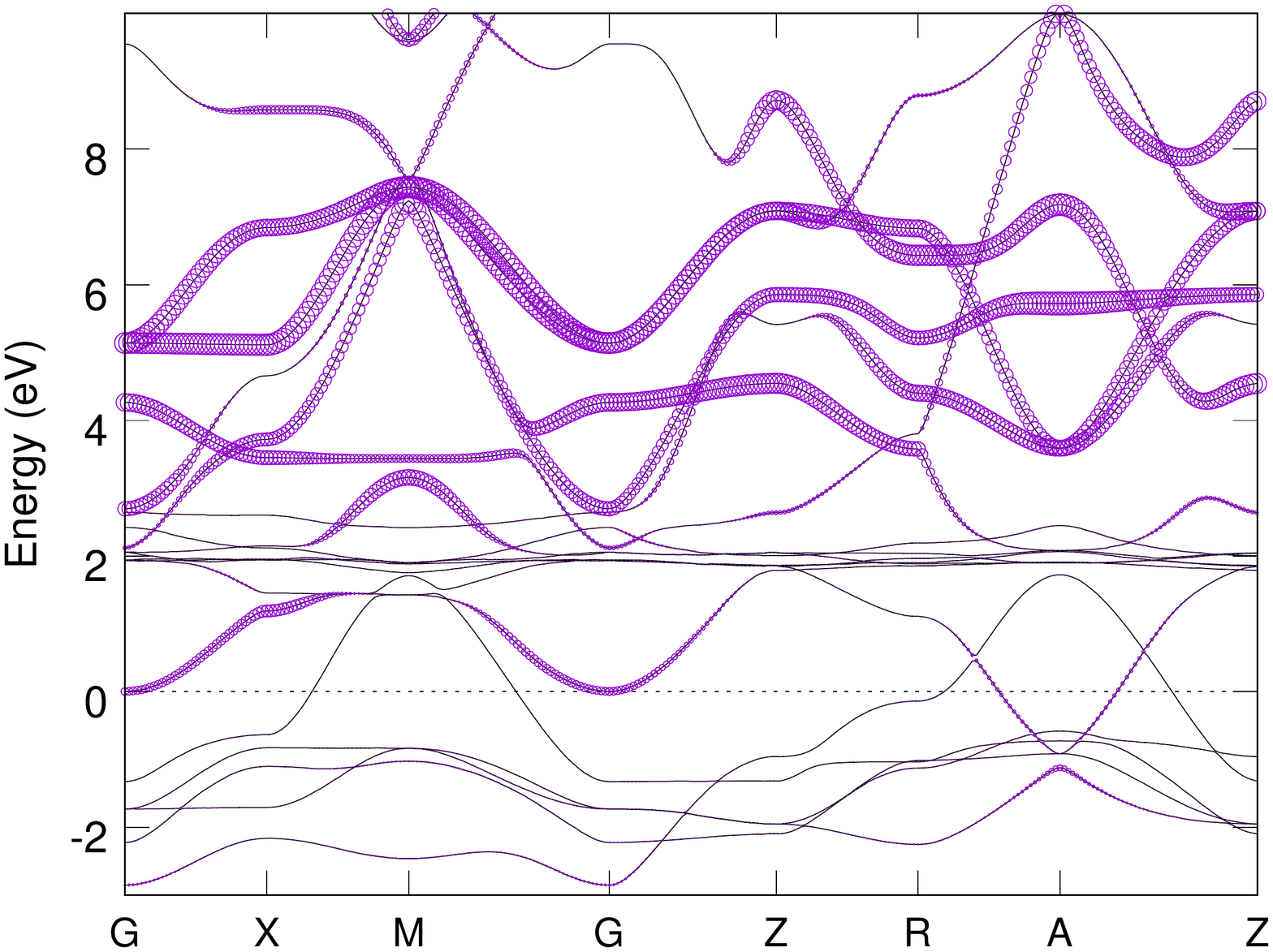}
    \caption{total La $d$}
  \end{subfigure}
  \begin{subfigure}[b]{0.49\columnwidth}
    \includegraphics[width=\linewidth]{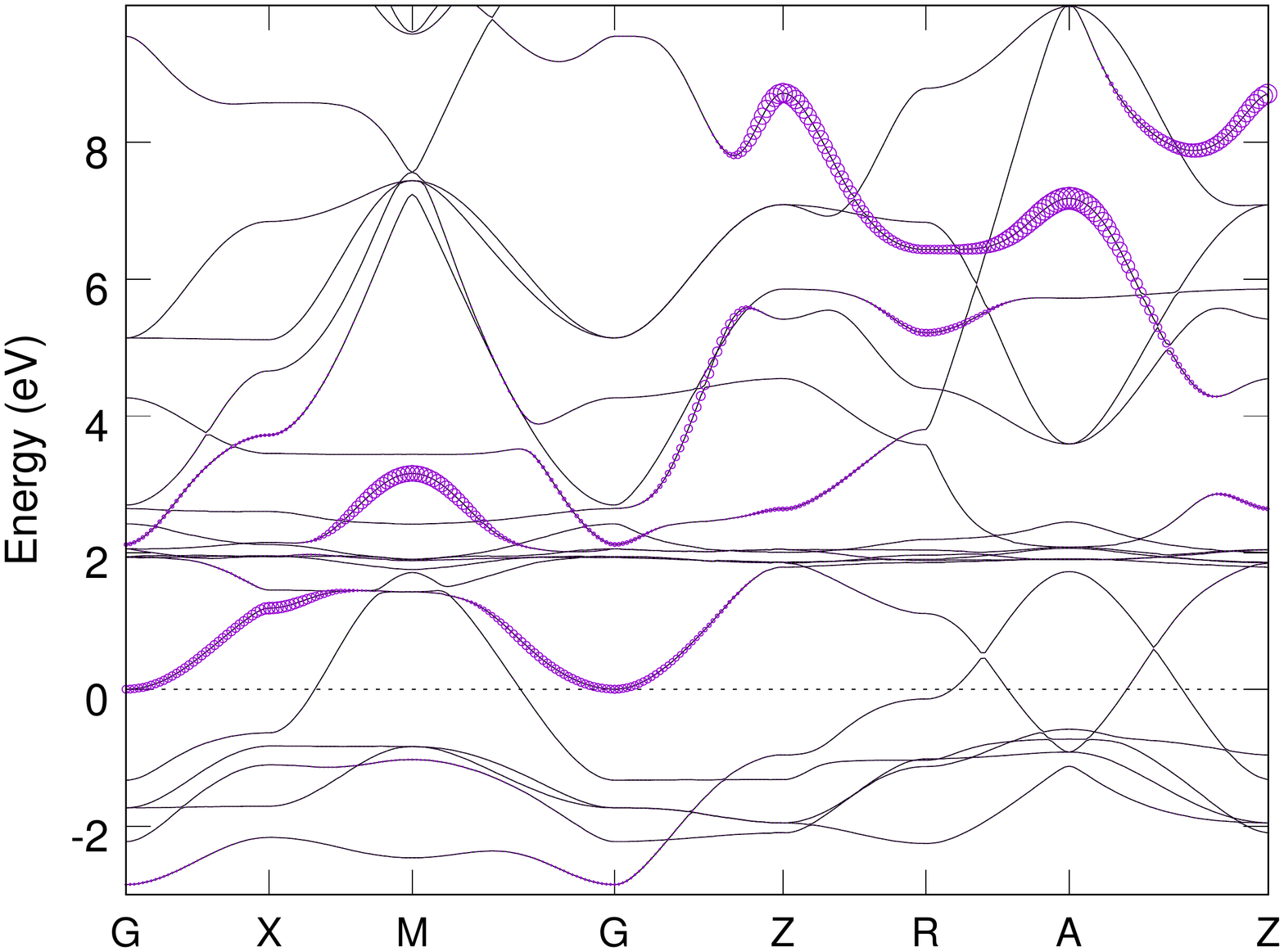}
    \caption{La $d_{3z^{2}-r^{2}}$}
  \end{subfigure}
  \begin{subfigure}[b]{0.49\columnwidth}
    \includegraphics[width=\linewidth]{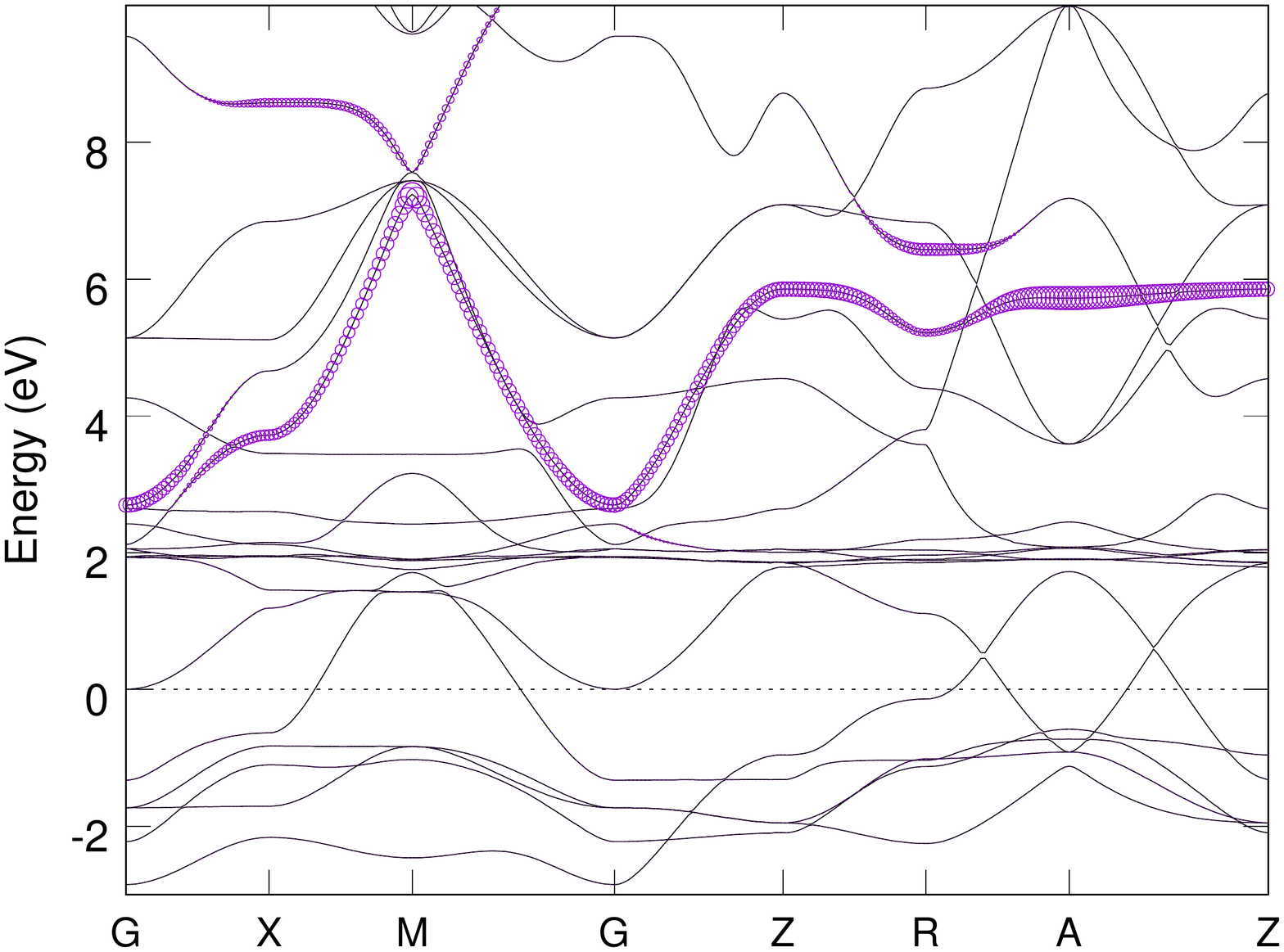}
    \caption{La $d_{x^{2}-y^{2}}$}
  \end{subfigure}
  \caption{Fat bands showing the amount of overlapping between Bloch states and La $d$ states and La $f$ states. (a) and (b): total $f$ and total $d$ character. (c) and (d): individual La $d_{3z^{2}-r^{2}}$ and La $d_{x^{2}-y^{2}}$ character.}
  \label{fat_bands_La}
\end{figure}

The La $d$-like and $f$-like bands are separately displayed in Fig.\ref{fat_bands_La}. From Fig.\ref{fat_bands_La} (a) and (b) we see the majority of La $d$-like bands are located above the $f$-like bands at +2 eV above the Fermi level. The break up of La $d_{3z^{2}-r^{2}}$ and $d_{x^{2}-y^{2}}$ in (c) and (d) help to clarify that the single band touching Fermi level from above at the $\Gamma$ point is clearly not of $d_{x^{2}-y^{2}}$ character. That band actually has limited La $d_{3z^{2}-r^{2}}$ character that is bound to $\Gamma$ point only, and its $d_{3z^{2}-r^{2}}$ character is not more significant than the amount of the mixing of Ni $d$ and O $p$ as shown in (a) and (b) in Fig.\ref{fat_bands_Ni}. Based on these observations we are not encouraged to include La $d_{3z^{2}-r^{2}}$ or La $d_{x^{2}-y^{2}}$ bands in the correlation window for model construction, though it had been done in other analysis \cite{PhysRevLett.125.077003}.

% --- figure 
% --- figure 

\subsection{Downfolding and Model Construction}

The single band crossing Fermi level is dominant Ni $d_{x^{2}-y^{2}}$-like that suggests a one-band model like that in the case of cuprate. The two-band $e_{g}$ model involving both the  $d_{x^{2}-y^{2}}$-like and $d_{3z^{2}-r^{2}}$-like bands is also motivated because of the entanglement of the two mentioned in the previous section. For a complete investigation of the relative correlation strength within the $d$ orbital subspace, we also consider the five-band model including all $d$-like bands. 

The downfolding technique is used to build the effective Hamiltonian in Wannier orbital basis. In the case of $LaNiO_{2}$ we have to deal with the situation of band entanglement, specifically the two $e_{g}$ bands are entangled with the other three Ni $d$-like bands and with the La $f$-like bands at around +2 eV above Fermi level.  

To handle such case, we employed the disentangle procedure introduced by T.Miyake and co-workers \cite{PhysRevB.80.155134}. First, a set of localized Wannier orbitals is constructed from a given correlation window (energy window around Fermi energy). It's large enough to include all or most weights of the target orbital character, so the bands of that orbital character can be well re-constructed in new Wannier orbital basis. The Wannier basis that yields the reconstructed bands (e.g. the two $e_{g}$ bands in the two-band model) will act as a subset of the basis, and we can call the spanned subspace the $d$-space. The basis for the rest of the entire Hilbert space, which we call it $r$-space, are then generated by Gram-Schmidt orthogonalization. By diagonalizing the Kohn-Sham Hamiltonian in the $r$-space, one gets a new set of eigen functions and eigen values. Namely, the Kohn-Sham Hamiltonian is re-diagonalized in the $d$-space and $r$-space separately, and the hybridization effect between the two subspaces is neglected. 

% ------ figure
\begin{figure}[H]
  \centering
  \includegraphics[width=0.9\columnwidth]{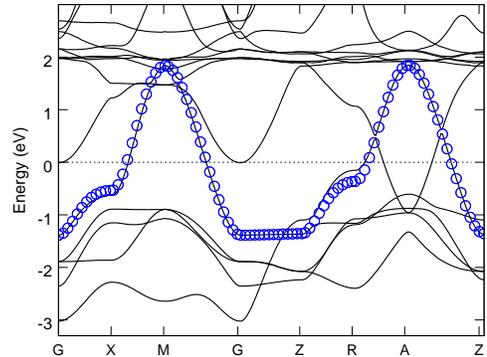}
  \caption{Downfolded and disentangled one-band for the $d_{x^{2}-y^{2}}$ subspace. Solid lines (black) are the original Bloch bands. Circles (blue) represents the reconstructed band in Wannier orbital basis. $E_{F}$ is at zero.} 
  \label{1band_model}
\end{figure}

% --- figure 
\begin{figure}[H]
  \centering
  \begin{subfigure}[b]{0.49\columnwidth}
    \includegraphics[width=\linewidth]{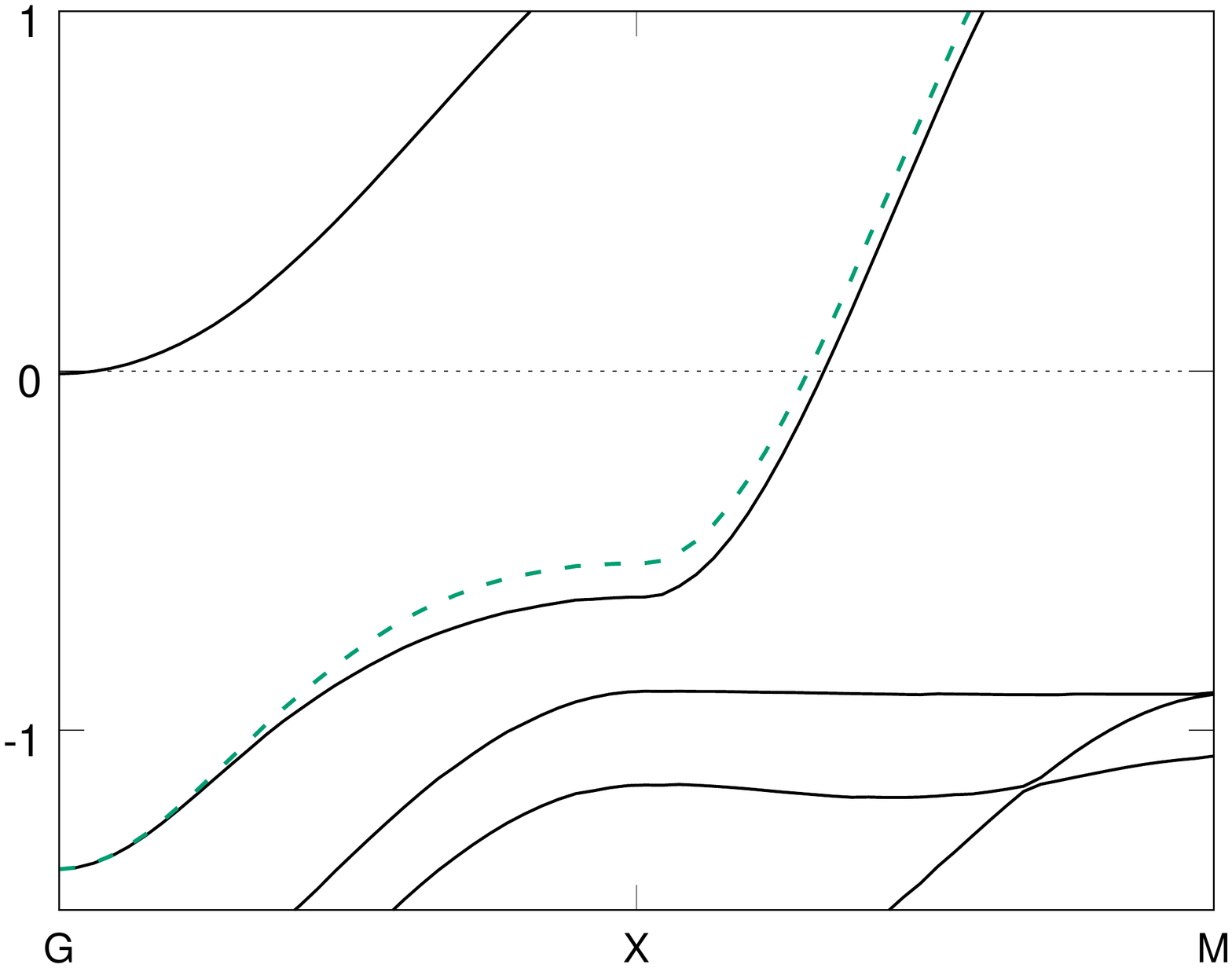}
     \caption{$d_{x^{2}-y^{2}}$}
  \end{subfigure}
  \begin{subfigure}[b]{0.49\columnwidth}
    \includegraphics[width=\linewidth]{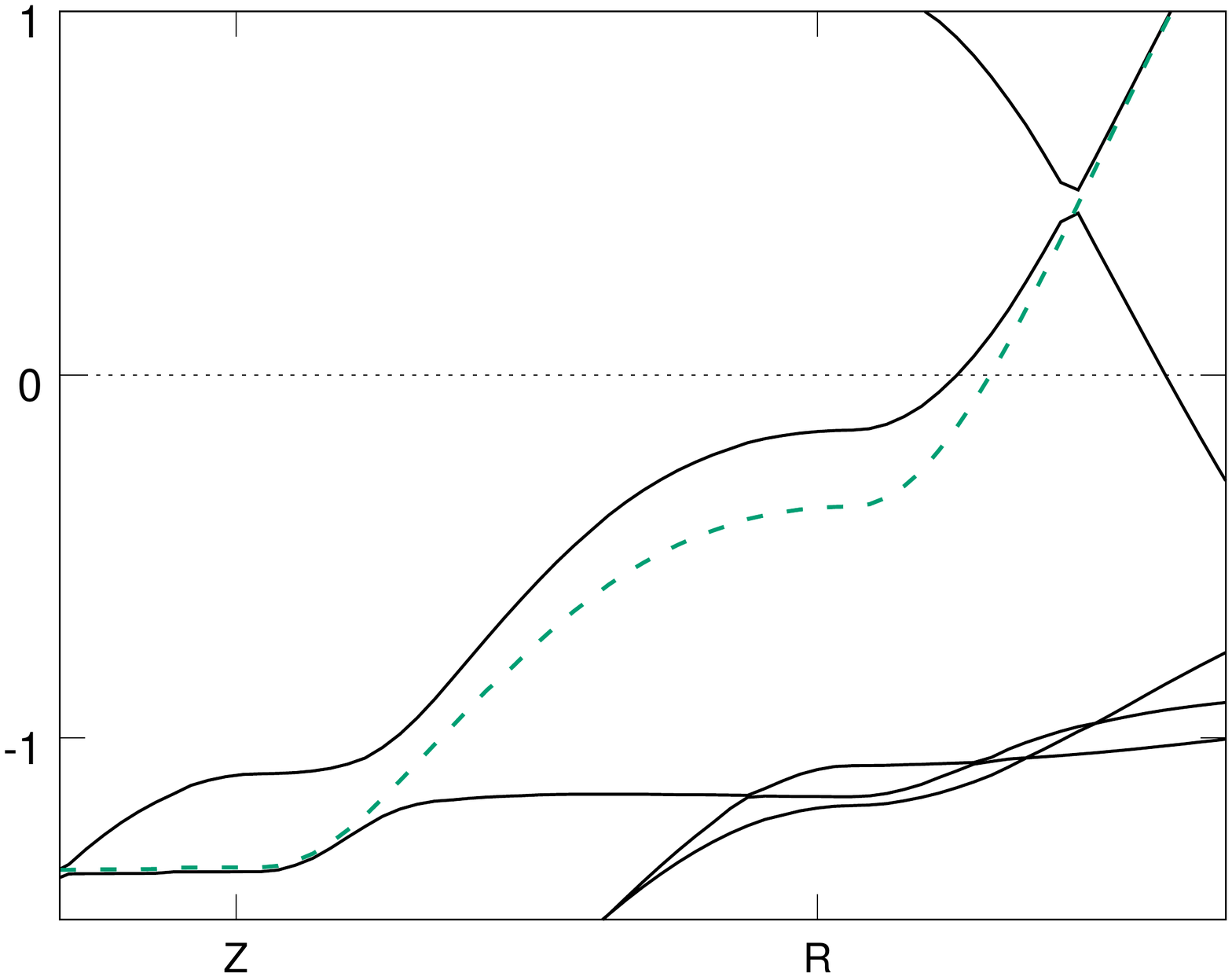}
    \caption{$d_{x^{2}-y^{2}}$}
  \end{subfigure}
  \begin{subfigure}[b]{0.49\columnwidth}
    \includegraphics[width=\linewidth]{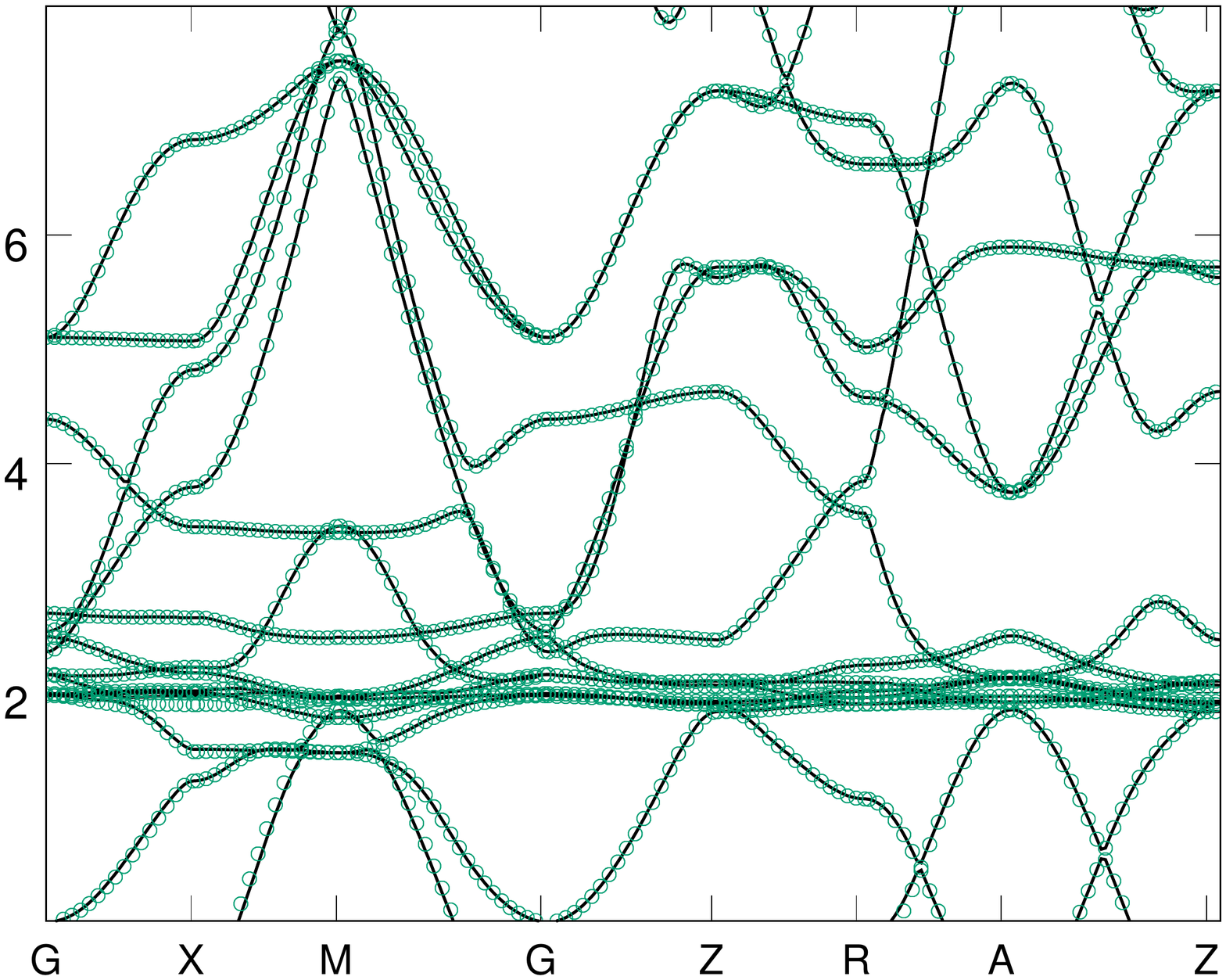}
    \caption{all reconstructed bands}
  \end{subfigure}
  \begin{subfigure}[b]{0.49\columnwidth}
    \includegraphics[width=\linewidth]{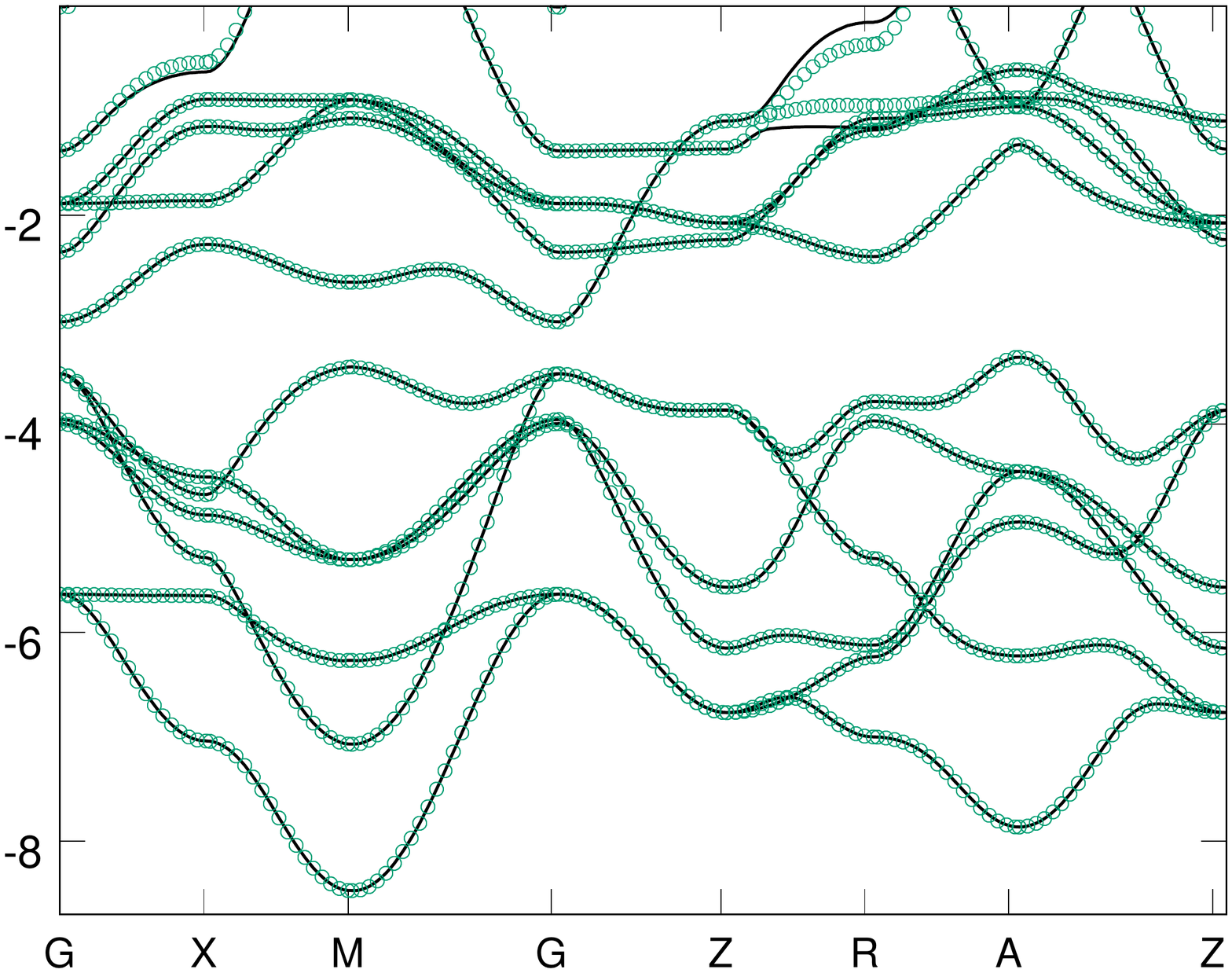}
    \caption{all reconstructed bands}
  \end{subfigure}
  \caption{Downfolded and disentangled bands for the one-band model. (a) and (b): enlarged details of the mismatching parts of the reconstructed $d_{x^{2}-y^{2}}$ band (green dash lines). (c) and (d): good matching of the reconstructed bands (green circles) in the $r$-space.}
  \label{1band_model_detail}
\end{figure}
% --- figure 

A good-working example of this disentangle method is the $fcc$ structure of Ni \cite{PhysRevB.80.155134}, where it's shown there can be small mismatching of bands within the $d$-space but all bands in the $r$-space are well reconstructed and match well. The similar effect is seen in our work, as shown in Fig.\ref{1band_model} and Fig.\ref{1band_model_detail} for the one-band model. Fig.\ref{1band_model_detail} (a) and (b) are zooming in the details around $k$=X, Z and R of Fig.\ref{1band_model}, where small band misalignment happens in the $d$-space. Fig.\ref{1band_model_detail} (c) and (d) show the good matching of the reconstructed bands in $r$-space in larger energy window (good matching is obtained throughout the entire energy spectrum).

\begin{figure}[H]
  \centering
  \includegraphics[width=0.9\columnwidth]{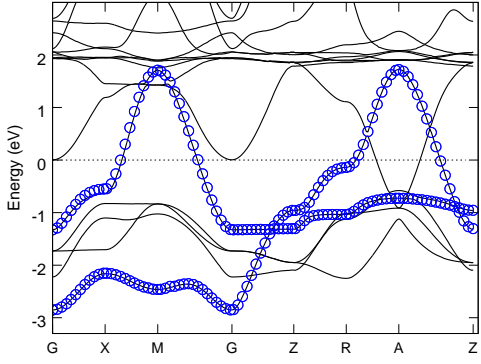}
  \caption{Downfolded and disentangled two-band model of the $d_{x^{2}-y^{2}}$ and $d_{3z^{2}-r^{2}}$ subspace of Ni. Solid lines (black) are the original Bloch bands. Circles (blue) represents the reconstructed $e_{g}$ bands in Wannier orbital basis. $E_{F}$ is at zero.} 
  \label{2band_model}
\end{figure}

\begin{figure}[H]
  \centering
  \includegraphics[width=0.9\columnwidth]{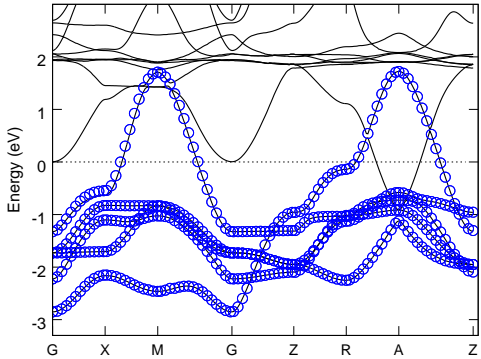}
  \caption{Downfolded and disentangled five-band model of all $d$ orbital subspace of Ni. Solid lines (black) are the original Bloch bands. Circles (blue) represents the reconstructed $d$ bands in Wannier orbital basis. $E_{F}$ is at zero.} 
  \label{5band_model}
\end{figure}

The same disentangled Wannier downfolding was done for the two-band model and the five-band model, as shown in Fig.\ref{2band_model} and Fig.\ref{5band_model} respectively. We noticed the misalignment around $k$=X, Z and R did not happen in this two model constructions. This is because both $e_{g}$ bands are included, thus the entanglement of the two, Fig.\ref{fat_bands_Ni} (c) and (d), is not cut. The result is good matching in both $d$-space and $r$-space. 

% ------ figure

% ------ figure

%\subsection{\label{sec:level2} cRPA calculation of U}
\subsection{cRPA calculation of U}

The fully quantum description of the Coulomb interaction between two electrons occupying a multi-orbital atomic site was derived and parameterized by Kanamori \cite{doi:10.1143/PTP.30.275}. Using the language of second quantization, the Coulomb interaction term of the Hamiltonian can be written as: 
\be
\begin{aligned}
\hat{H}_{int} & = \mathcal{U} \sum_{l} \hat{n}_{l\sigma} \hat{n}_{l\bar{\sigma}} \\ 
& + \frac{1}{2} \sum_{l \neq l'}\sum_{\sigma} [\mathcal{U}' \hat{n}_{l\sigma} \hat{n}_{l'\bar{\sigma}} + (\mathcal{U}'-\mathcal{J}) \hat{n}_{l\sigma} \hat{n}_{l'\sigma} ] \\
& + \frac{1}{2} \sum_{l \neq l'}\sum_{\sigma} [\mathcal{J} \hat{c}^{\dagger}_{l\sigma} \hat{c}^{\dagger}_{l'\bar{\sigma}} \hat{c}_{l\bar{\sigma}} \hat{c}_{l'\sigma} + \mathcal{J}_{c} \hat{c}^{\dagger}_{l\sigma} \hat{c}^{\dagger}_{l\bar{\sigma}} \hat{c}_{l'\bar{\sigma}} \hat{c}_{l'\sigma} ]
\end{aligned}
\label{eq:00}
\ee
where $l$ and $l'$ are angular momentum quantum numbers, $\sigma$ and $\bar{\sigma}$ are spin and opposite-spin quantum numbers, $\hat{c}^{\dagger}_{l\sigma}$ and $\hat{c}_{l\sigma}$ are electron creation and annihilation operators for state $(l,\sigma)$, and $\hat{n}_{l\sigma} \equiv \hat{c}^{\dagger}_{l\sigma} \hat{c}_{l\sigma}$ is the density operator.

The first line of Eq.(\ref{eq:00}) represents two electrons occupying the same orbital (must be opposite spins due to the Pauli principle). The second line includes the situations where two electrons occupying two different orbitals, and they can be same or opposite spins. The two terms in the third line are called the spin-flip (coefficient $\mathcal{J}$) and pair-hopping (coefficient $\mathcal{J}_{c}$), which cannot be written as density-density interaction form and are often neglected (at least for nonmagnetic systems spin-flip and pair-hopping processes should have little influence). Thus, for most common cases, quantum description of the on-site Coulomb interaction requires the calculation of $\mathcal{U}$, $\mathcal{U}'$ and $\mathcal{J}$, which are the quantities we calculate in the current work. 

One way to calculate the Coulomb interaction U from first principles is the constrained Random Phase Approximation (cRPA), that has been well explained in the literature \cite{PhysRevB_frequ_depend_U,U_from_cRPA}. In this section we briefly go over the original idea, followed by description of the implementation based on the density response function. The resulting U matrices from the three models are presented at the end. 

The cRPA calculation is based on the RPA approximation where the constrain means excluding a group of orbitals to get a reduced polarization function. By doing that one gets an estimation of the partially screened Coulomb interaction for a selected group of bands of interest, e.g. localized $d$ orbitals of a transition metal atom. The original RPA approximation considers the particle-hole polarization between all possible pairs of occupied state and unoccupied state. Within DFT the particle-hole polarization can be expressed as \cite{PhysRevLett.76.1212}:
\be
\begin{aligned}
P(\bm{r},\bm{r}',\omega) & = \sum_{i}^{occ.} \sum_{j}^{unocc.} [\psi_{i}^{*}(\bm{r}) \cdot \psi_{j}(\bm{r}\ensuremath{'}) \cdot \psi_{j}^{*}(\bm{r}) \cdot \psi_{i}(\bm{r}\ensuremath{'})] \\
& \times ( \frac{1}{ \omega - \varepsilon_{j} + \varepsilon_{i} + i\delta} + \frac{1}{ \omega + \varepsilon_{j} + \varepsilon_{i} - i\delta} ) 
\end{aligned}
\label{eq:01}
\ee
\noindent where $\psi_{i}$ and $\varepsilon_{i}$ are the eigen functions and eigen energies of the Kohn-Sham Hamiltonian. The summation over $i$ and $j$ is restricted such that $i$ is an occupied state and $j$ is an unoccupied state. 

Within DFT, the chosen correlation window contains the selected bands of interest that have a particular orbital character, e.g. the $d$-like bands of Ni in our case. Since the previous section we have followed the convention in the literatures where we labelled the bands of interests as the $d$-space and the bands outside the correlation window as the $r$-space. If both the occupied state and the unoccupied state are within the $d$-space, then the polarization contributes to $P_{d}(r,r';\omega)$. All the other pairs of occupied and unoccupied states contribute to $P_{r}$. Thus, the total polarization is divided into two parts: $P = P_{d} + P_{r}$. The $P_{r}$ is the quantity related to the partially screened Coulomb interaction \cite{PhysRevB.70.195104}: 
\be
W_{r}(\bm{r},\bm{r}',\omega)=[1 - v \cdot P_{r}(\bm{r},\bm{r}',\omega)]^{-1} \cdot v 
\label{eq:02}
\ee
\noindent where $v$ is the bare Coulomb interaction. 

According to the Hedin’s equations and the GW approximation, the total polarization, $P$, screens the bare Coulomb interaction, $v$, to give the fully screened interaction $W$, namely: 
\be
W(\bm{r},\bm{r}',\omega)=[1-\nu \cdot P(\bm{r},\bm{r}',\omega)]^{-1} \cdot \nu 
\label{eq:02b}
\ee
Eq.(\ref{eq:02}) follows similar interpretation as Eq.(\ref{eq:02b}) where $P_{r}$ screens the bare Coulomb interaction to give the partially screened interaction $W_{r}$. At last the $U(\omega)$ matrices is calculated from the screened Coulomb interaction $W_{r}$ \cite{5644911}: 
\be
U_{nn'}^{TT'}(\omega) \equiv \iint |w_{n}^{\bm{T}}(\bm{r})|^{2} W_{r}(\bm{r},\bm{r}',\omega) |w_{n'}^{\bm{T}'}(\bm{r}')|^{2} d\bm{r} d\bm{r}' 
\label{eq:03}
\ee
where $w_{n}(\bm{r})$ is the $n$th Wannier orbital within the $d$-space. The subscript $n$ has same feature as the angular momentum quantum number $l$ in Eq.(\ref{eq:00}). The superscript $T$ and $T'$ are real space lattice vectors, indicating the location of the Wannier center in real space lattice.  

The key quantity in realization of the above described cRPA is to calculate the polarization function $P$, which is the density response function $\chi$. In general the Kohn-Sham density response function $\chi^{KS}$ is related to the general response function $\chi$ through the following integral equation \cite{PhysRevLett.76.1212}: 
\be
\begin{aligned}
\chi(\bm{r},\bm{r}',\omega) &= \chi^{KS}(\bm{r},\bm{r}',\omega) + \iint d\bm{r}_{1}d\bm{r}_{2} [ \chi^{KS}(\bm{r},\bm{r}_{1};\omega) \\
& \cdot \left( \frac{1}{|\bm{r}_{1}-\bm{r}_{2}|}+f^{xc}(\bm{r}_{1},\bm{r}_{2};\omega)\right) \cdot \chi(\bm{r}_{2},r';\omega) ]
\end{aligned}
\label{eq:04}
\ee
\noindent where the Kohn-Sham response function can be written as \cite{PhysRevLett.76.1212,5644911}: 
\be
\begin{aligned}
\chi^{KS}(r,r\ensuremath{'};\omega) = \sum_{i,j} \frac{(f_{i}-f_{j})\psi_{i}^{*}(r)\psi_{j}(r\ensuremath{'})\psi_{j}^{*}(r)\psi_{i}(r\ensuremath{'})}{\omega-\varepsilon_{j}+\varepsilon_{i} +i\delta}
\end{aligned}
\label{eq:05}
\ee
\noindent where $f_{i}$ and $\varepsilon_{i}$ are the occupancy and eigen energy of the eigen state $\psi_{i}$, and $f^{xc}$ is the functional derivative of the exchange-correlation potential with respect to the charge density which is neglected in the random phase approximation \cite{PhysRevLett.76.1212} if we assume non-interacting electrons.

The constrain (excluding the contribution from $d$-space) is directly applied to $\chi^{KS}$ to get $\chi_{r}^{KS}$. Then $\chi_{r}$ is solved from the \textit{reduced} version of Eq.(\ref{eq:04}):
\be
\begin{aligned}
\chi_{r}(\bm{r},\bm{r}',\omega) &= \chi_{r}^{KS}(\bm{r},\bm{r}',\omega) + \iint d\bm{r}_{1}d\bm{r}_{2} [ \chi_{r}^{KS}(\bm{r},\bm{r}_{1};\omega) \\
& \cdot \left( \frac{1}{|\bm{r}_{1}-\bm{r}_{2}|}+f^{xc}(\bm{r}_{1},\bm{r}_{2};\omega)\right) \cdot \chi_{r}(\bm{r}_{2},r';\omega) ]
\end{aligned}
\label{eq:04b}
\ee
The rest steps is based on the linear response theory \cite{Dyachenko2012}, where the partially screened Coulomb interaction $W_{r}$ is related to inverse dielectric function $\varepsilon^{-1}$ and bare Coulomb interaction $v$ and $\varepsilon^{-1}$ can be obtained from $\chi_{r}$: 
\be
\begin{aligned}
W_{r}(\bm{r}_{1},\bm{r}_{2},\omega) &= \int dr [\varepsilon^{-1}(\bm{r}_{1},\bm{r},\omega) \cdot \nu(\bm{r},\bm{r}_{2})] \\
&= \int dr [\left( 1+\nu\cdot\chi_{r}(\bm{r}_{1},\bm{r},\omega) \right) \cdot \nu(\bm{r},\bm{r}_{2})]
\end{aligned}
\label{eq:06}
\ee
At last one uses Eq.(\ref{eq:03}) to get $U^{TT'}_{nn'}(\omega)$.

%In practise the above equations are solved in their Fourier transformed forms:
%\be
%\begin{aligned}
%\chi_{GG'}(q,\omega) &= \chi^{KS}_{GG'}(q,\omega) + 
%\sum_{G_{1}G_{2}}[ \chi^{KS}_{GG_{1}}(q,\omega) \cdot ( v_{G_{1}+q}\delta_{G_{1}G_{2}} \\
%& + f_{G_{1}G_{2}}^{xc}(q,\omega)) \cdot \chi_{G_{2}G'}(q,\omega) ]
%\end{aligned}
%\label{eq:05}
%\ee
%\noindent where $q$ is reciprocal lattice vector within the first Brillouin zone and $v_{G+q}=4\pi /|G+q|^{2}$ is the expansion coefficient of the bare Coulomb interaction. $f^{xc}$ is the functional derivative of the exchange-correlation potential with respect to the charge density which is neglected in the random phase approximation \cite{PhysRevLett.76.1212}.
%
%\be
%W_{r,GG'}(q,\omega)=v_{G+q}\delta_{GG'} + v_{G+q} \cdot \chi_{r,GG'}(q,\omega) \cdot v_{G'+q}
%\label{eq:09}
%\ee
The above described calculations have been implemented in the Exciting-Plus code (a modified version of ELK code) \cite{ELK_code,5644911}. In practise, 100 empty bands are included in the ground state calculation. We have benchmarked the method using late transition monoxides NiO, CoO, FeO and MnO and got results in agreement with other implementations of essentially the same method \cite{PhysRevB.100.035104}. The one, two and five Ni $d$ bands are excluded in calculating the $\chi_{r}^{KS}$, for the three models respectively. The resulting $U$ matrices are within the Kanamori parameterization described at the beginning of this section and the parameters $\mathcal{U}$, $\mathcal{U}'$ and $\mathcal{J}$ are often organized as a $U$ matrix and a $J$ matrix, respectively, for example for the two-band model of $d_{3z^{2}-r^{2}}$ and $d_{x^{2}-y^{2}}$ subspace: 
\[
\begin{blockarray}{ccc}
d_{3z^{2}-r^{2}} & d_{x^{2}-y^{2}} \\
\begin{block}{(cc)c}
\mathcal{U}  &  \mathcal{U}'  &  d_{3z^{2}-r^{2}}  \\
\mathcal{U}' &  \mathcal{U}   &  d_{x^{2}-y^{2}}   \\
\end{block}
\end{blockarray}
\qquad
\begin{blockarray}{ccc}
d_{3z^{2}-r^{2}} & d_{x^{2}-y^{2}} \\
\begin{block}{(cc)c}
\mathcal{U}  &  \mathcal{J}   &  d_{3z^{2}-r^{2}}  \\
\mathcal{J}  &  \mathcal{U}   &  d_{x^{2}-y^{2}}   \\
\end{block}
\end{blockarray}
\]

An important feature of the cRPA calculation is that it actually does not restrict to only on-site interaction because the key quantity $W_{r}$ is not a local function, though most usages of the method focus on on-site interaction only, i.e. $T=T'$ in Eq.(\ref{eq:03}). If the Wannier orbital centers are chosen to be on the neighbor sites, e.g. $T - T' = N_{1}\cdot\bm{R}_{1} + N_{2}\cdot\bm{R}_{2} + N_{3}\cdot\bm{R}_{3}$ with $\bm{R}_{1},\bm{R}_{2},\bm{R}_{3}$ being the three lattice vectors of the primitive cell and $N_{1},N_{2},N_{3}$ being integers, then the physical meaning of the calculated $U(\omega)$ would be the Coulomb interaction when two electrons occupy two neighbor Ni sites. The interaction strength would be of course smaller than the on-site value, but quantitative values of $U^{n.n.}(\omega)$ and $U^{n.n.n.}(\omega)$ are important parameters and are desirable especially when studying non-local correlations.  

%===================================================
\section{on-site and nearest neighbor U}

\begin{figure}[H]
  \centering
  \includegraphics[width=0.9\columnwidth]{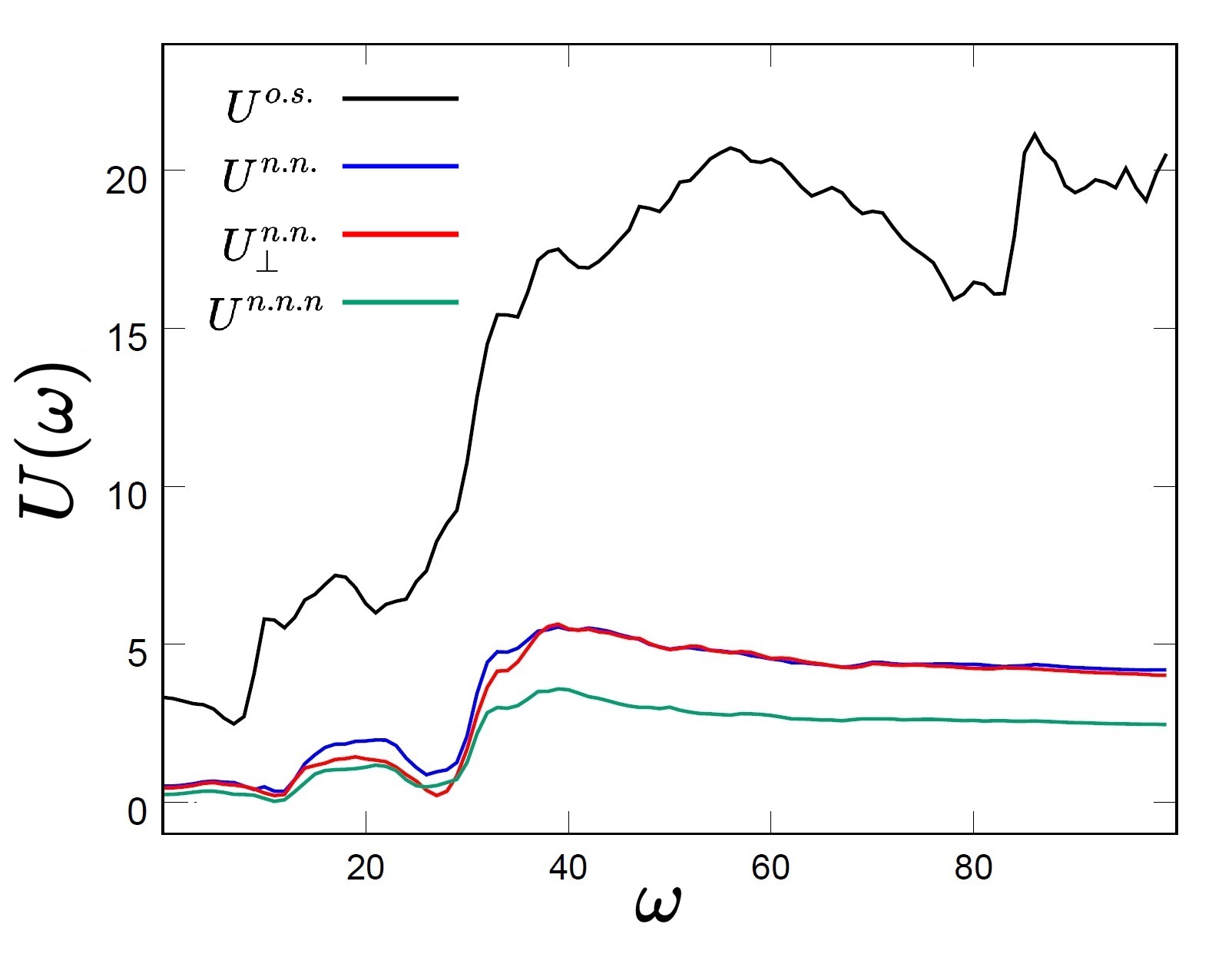}
  \caption{Frequency dependent $U$ of the one-band model. $U(\omega)$ is in unit of eV.} 
  \label{U_1band_model}
\end{figure}

We have calculated the following Coulomb interaction matrices as labelled in Fig.\ref{crystal_LaNiO2}:
the in Ni-O plane interactions $U^{o.s.}(\omega)$, $U^{n.n.}(\omega)$, $U^{n.n.n.}(\omega)$ and the inter Ni-O plane interaction $U_{\perp}^{n.n.}(\omega)$, where $o.s.$ stands for on-site, $n.n.$ stands for nearest neighbor, and $n.n.n.$ stands for next nearest neighbor. 
From the one-band ($d_{x^{2}-y^{2}}$) model:
\[ 
U^{o.s.}(\omega=0) = J^{o.s.}(\omega=0) = 
\left( 
\begin{array}{c}
\mathcal{U}
\end{array} 
\right) = 
\left( 
\begin{array}{c}
3.32 
\end{array} 
\right)
\] 
\[ 
U^{n.n.}(\omega=0) = J^{n.n.}(\omega=0) = 
\left( 
\begin{array}{c}
\mathcal{U}
\end{array} 
\right) = 
\left( 
\begin{array}{c}
0.50 
\end{array} 
\right)
\] 
\[ 
U^{n.n.}_{\perp}(\omega=0) = J^{n.n.}_{\perp}(\omega=0) = 
\left( 
\begin{array}{c}
\mathcal{U}
\end{array} 
\right) = 
\left( 
\begin{array}{c}
0.44 
\end{array} 
\right)
\] 
\[
U^{n.n.n.}(\omega=0) = J^{n.n.n.}(\omega=0) = 
\left( 
\begin{array}{c}
\mathcal{U}
\end{array} 
\right) = 
\left( 
\begin{array}{c}
0.24 
\end{array} 
\right)
\]
All the above values are in unit of eV. The on-site interaction is 3.32 eV for two electrons sitting on the single $d_{x^{2}-y^{2}}$ orbital, while the non-on-site values correspond to the situations where one electron is in one $d_{x^{2}-y^{2}}$ orbital and another electron is in a different site $d_{x^{2}-y^{2}}$ orbital. 
The frequency dependency of the above listed quantities are plot in Fig.(\ref{U_1band_model}). The ratio of $U^{o.s.}/U^{n.n.}$ increase from 6.6 at $\omega=0$ to about 4 as $\omega$ increases to $\omega>40$. Both $U^{n.n.}$ and $U^{n.n.}_{\perp}$ become almost $\omega$-independent and have identical value of about 5 eV when $\omega>40$, that suggests the inter Ni-O layer interaction $U^{n.n.}_{\perp}$ in on equal footing with the $U^{n.n.}$ within Ni-O plane. 

% ------ figure

For the two-band ($d_{3z^{2}-r^{2}}$ and $d_{x^{2}-y^{2}}$) model, we got the following U and J matrices at $\omega=0$: 

\[ 
U^{o.s.} = 
%\left( \begin{array}{cc}
%\mathcal{U}  & \mathcal{U}'  \\
%\mathcal{U}' & \mathcal{U}   \end{array} \right) =
\left( \begin{array}{cc}
2.86 & 1.79  \\
1.79 & 3.27  \end{array} \right); \qquad
J^{o.s.} = 
%\left( \begin{array}{cc}
%\mathcal{U}  & \mathcal{J}   \\
%\mathcal{J}  & \mathcal{U}   \end{array} \right) =
\left( \begin{array}{cc}
2.86 & 0.56  \\
0.56 & 3.27  \end{array} \right);
\] 

\[ 
U^{n.n.} = 
%\left( \begin{array}{cc}
%\mathcal{U}  & \mathcal{U}'  \\
%\mathcal{U}' & \mathcal{U}   \end{array} \right) =
\left( \begin{array}{cc}
0.36 & 0.41  \\
0.41 & 0.48  \end{array} \right); \qquad
J^{n.n.} = 
%\left( \begin{array}{cc}
%\mathcal{U}  & \mathcal{J}   \\
%\mathcal{J}  & \mathcal{U}   \end{array} \right) =
\left( \begin{array}{cc}
0.36 & 0.01  \\
0.01 & 0.48  \end{array} \right);
\]

\[ 
U^{n.n.}_{\perp} = 
%\left( \begin{array}{cc}
%\mathcal{U}  & \mathcal{U}'  \\
%\mathcal{U}' & \mathcal{U}   \end{array} \right) =
\left( \begin{array}{cc}
0.63 & 0.49  \\
0.49 & 0.41  \end{array} \right); \qquad
J^{n.n.}_{\perp} = 
%\left( \begin{array}{cc}
%\mathcal{U}  & \mathcal{J}   \\
%\mathcal{J}  & \mathcal{U}   \end{array} \right) =
\left( \begin{array}{cc}
0.63 & 0.00  \\
0.00 & 0.41  \end{array} \right);
\]

\[ 
U^{n.n.n.} = 
%\left( \begin{array}{cc}
%\mathcal{U}  & \mathcal{U}'  \\
%\mathcal{U}' & \mathcal{U}   \end{array} \right) =
\left( \begin{array}{cc}
0.22 & 0.22  \\
0.22 & 0.22  \end{array} \right); \qquad
J^{n.n.n.} = 
%\left( \begin{array}{cc}
%\mathcal{U}  & \mathcal{J}   \\
%\mathcal{J}  & \mathcal{U}   \end{array} \right) =
\left( \begin{array}{cc}
0.22 & 0.00  \\
0.00 & 0.22  \end{array} \right);
\]

% ------ figure
\begin{figure}[H]
  \centering
  \includegraphics[width=0.9\columnwidth]{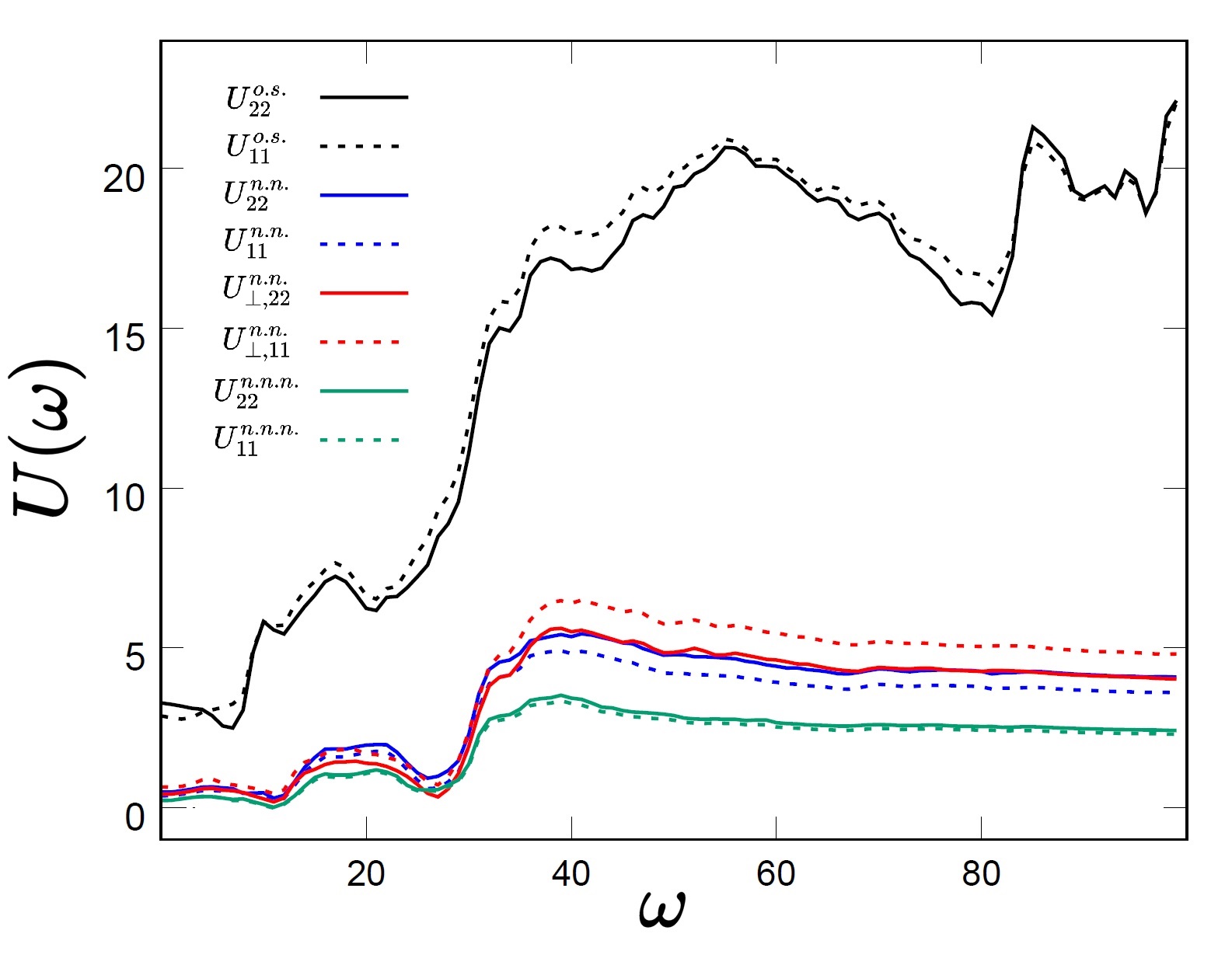}
  \caption{Frequency dependency of the diagonal element of the $U$ matrix of the two-band model. Unit is eV.} 
  \label{U_2band_model}
\end{figure}

It's clear to see from Fig.(\ref{U_2band_model}) the two diagonal elements of on-site interaction (black solid and black dash curves) have almost identical frequency dependency over the whole range, though they differ by about 0.4 eV at $\omega=0$. It suggests the Coulomb interaction strength is same no matter two electrons both on the $d_{x^{2}-y^{2}}$ orbital or both on the $d_{3z^{2}-r^{2}}$ orbital. And the observation is in consistency with that from the one-band model. Another observation that's same as the one-band model is the almost identical in-plane and inter-plane nearest neighbor interactions, $U^{n.n.}_{22}$ and $U^{n.n.}_{\perp,22}$ for the $d_{x^{2}-y^{2}}$ orbital (blue and red solid lines). However the $U^{n.n.}_{11}$ and $U^{n.n.}_{\perp,11}$ for the $d_{3z^{2}-r^{2}}$ orbital (blue and red dash lines) are different, where the inter Ni-O plane interaction is greater. When $\omega>40$ the difference keeps at about 1.25 eV steady. An additional results from the two-band model is the off-diagonal elements of $J^{n.n.}$ and $J^{n.n.}_{\perp}$ being zero, i.e. the parameter $\mathcal{J}$ is zero. Recall the two density-density terms in the second line of Eq.(\ref{eq:00}) and the picture where one electron sits in the $d_{x^{2}-y^{2}}$ (or $d_{3z^{2}-r^{2}}$) orbital and another electron sits in the other orbital of a nearest neighbor site, $\mathcal{J}=0$ means the interaction strength does not care whether the two spins are parallel or anti-parallel. And, the same for $J^{n.n.n.}$. 

At last, $U^{n.n.n.}$ is of same features in the two models. The two-band model does not present any new feature for the $d_{3z^{2}-r^{2}}$ orbital, it's same as the $d_{x^{2}-y^{2}}$ orbital. 

% ------ figure
\begin{figure}[H]
  \centering
  \includegraphics[width=0.9\columnwidth]{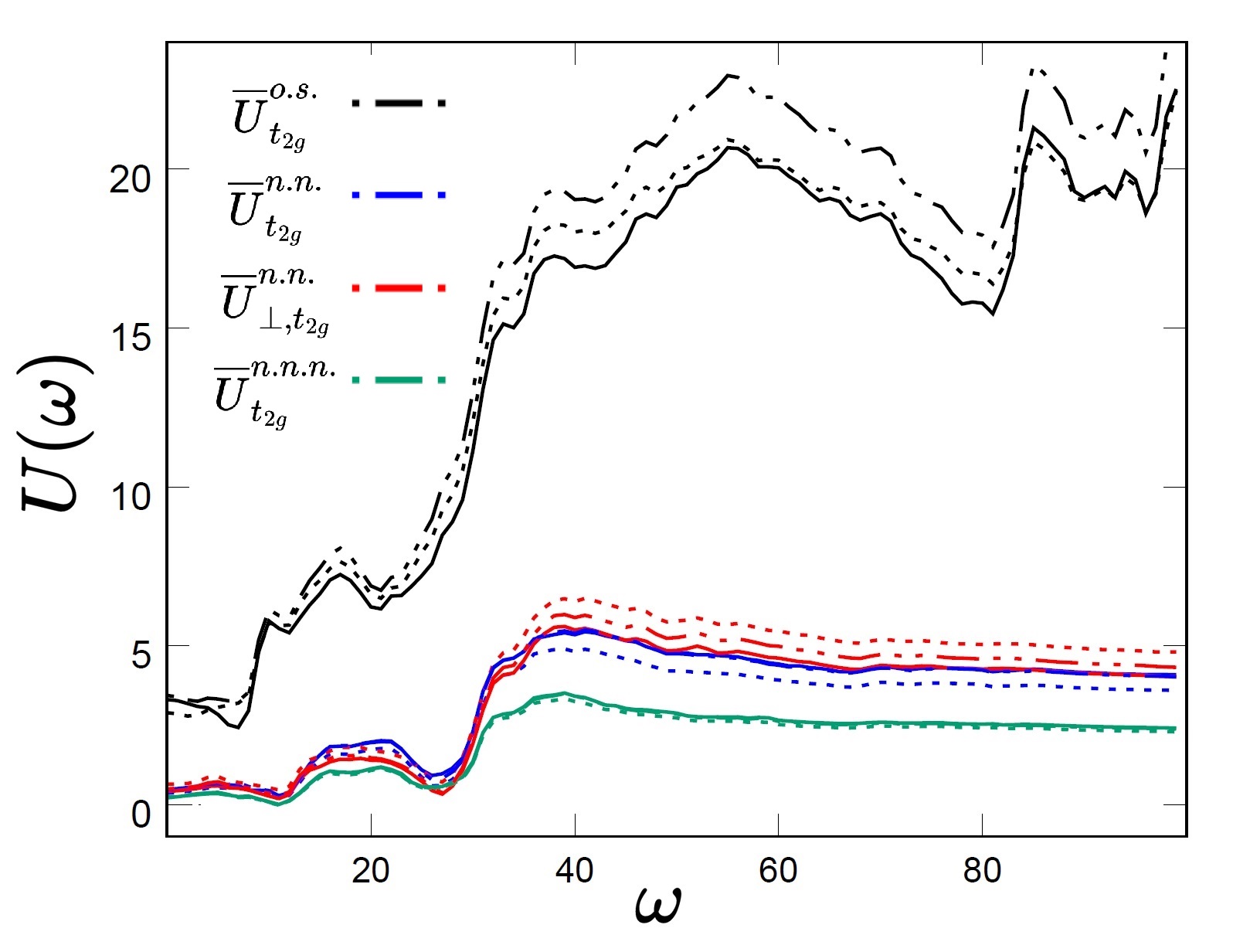}
  \caption{Frequency dependency of the diagonal element of the $U$ matrix of the five-band model. Unit is eV.} 
  \label{U_5band_model}
\end{figure}

\section{Summary and Conclusion}

In summary, we performed DFT calculation of bulk $LaNiO_{2}$ in its non-magnetic phase, and constructed different model Hamiltonians including the single $d_{x^{2}-y^{2}}$ orbital, the $e_{g}$ orbitals and all five $d$ orbitals of Ni. 
And we performed cRPA calculations of the Coulomb interactions for the on-site $d$ orbitals and Coulomb interactions for the $d$ orbitals between neighbor sites, for the constructed models. 
The resulting Coulomb interaction parameters of the $d_{x^{2}-y^{2}}$ orbital are consistent within all three models. The results for the $e_{g}$ orbitals from the two-band model agree with that from the five-band model too. 

The ratio of $U^{o.s.}(\omega)/U^{n.n.}(\omega)$ is found to be 6.6 at $\omega=0$ and drops to about 4 when $\omega>40$, indicating a pretty strong non-local Coulomb interaction. The inter Ni-O plane nearest neighbor Coulomb interaction is found to be almost exactly same as the in plane one, $U^{n.n.}(\omega)/U_{\perp}^{n.n.}(\omega) \approx 1$. It suggest the material is non-locally correlated in all x, y and z directions. In the longer range, we found $U^{n.n.n.}$ is about 50\%-60\% of the $U^{n.n.}$ over the entire frequency range. 

In conclusion, the presented work provides a quantitative and detailed description of the local and non-local Coulomb interaction strengths of the Ni $d$-orbitals in $LaNiO_{2}$ from first principle. The numerical study suggests a not very strong but very non-local electron correlation of this material, that could benefit future DFT calculations as well as model calculations of the nickelate family of superconductors. 

\clearpage
\bibliography{main}
\end{document}